Authors: E. Simoncini (1), A. Kleidon (1), E. Gallori (2)
((1) Max Planck Institute für Biogeochemie, Jena, Deutschland. (2) Dipartimento di Fisica e Astronomia, Università di Firenze, Italia)

In this paper we demonstrate how chemical free energy can be produced by a geological process. We provide a thermodynamic framework in which to assess how life emerged at the off-axis hydrothermal vent system; the RNA - clays system has been investigated from the entropic point of view, showing that the stabilization of the system in a state further away from equilibrium state, by an inorganic heterogeneous compartmetalization phenomena, is able to produce chemical free energy useful for RNA self - replication.

# Thermodynamics of chemical free energy generation in off-axis hydrothermal vent systems and its consequences for compartmentalization and the emergence of life

## 1. Introduction: how does Disequilibrium work on Earth surface

The Earth system is maintained far from thermodynamic equilibrium by a continuous flow of energy from the sun and from interior matter cooling and nuclear decays. This disequilibrium condition forces the Earth to dissipate energy, and its associated entropy production is a measure of the degree of Earth system irreversibility (Kleidon 2009). The sequestration of chemical components from an once homogeneous system into separate domains (chemical differentiation) results in a loss of entropy, which is possible because differentiation occurs in conjunction with a greater production of entropy associated with the decay of thermal gradients by various forms of heat flow (Prigogine 1967. Rosing et al., 2006).

When a system is pushed far from the equilibrium, complex processes rise, with feedbacks (Prigogine, 1967; 1978). While negative feedbacks generate in steady states, positive ones generate during periods of rapid evolution. Hence, from Prigogine theory on Dissipative Structures descends tat the development of structures that maintain a disequilibrium state for a longer time is thermodynamically favored in a far from equilibrium system. The constraints for this process are the First and Second Principles of thermodynamics, so the free energy balance throughout this open system (Schrödinger 1944).

Compartmentalization and the confinement of unstable compounds is one of the most important examples of this process. A chemically "unstable" compound reacts easily, hence a far from equilibrium system is capable of storing chemical energy. Nature is full of examples in which a new boundary is built in order to take more energy from an already existing process. A sketch of the theoretical framework exposed in this paper is shown in Figure 1.

Figure1: Overview of the concept presented in this paper. Chemiosmosis, complex chemical compounds and proto‑metabolism are emergent properties of disequilibrium in the Earth system.

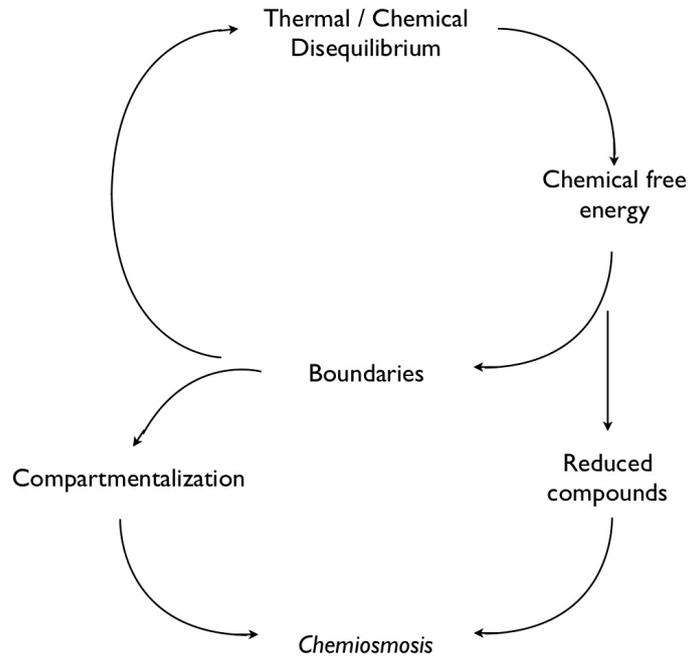

In general, when a *boundary* is built, it cuts down on the diffusive loss and thereby enhances the ability and efficiency to utilize free energy. Boundaries delimit an open thermodynamic system from its environment, and act as an exchange medium, which, hence, creates selectivity for chemical species throughout different diffusion. Compartmentalization is needed for a Darwinian evolution (Hanczyc et al., 2003), and an already existing dissipative environment is required (Corliss 1990).
For molecular complexity to evolve in the ancient Earth oceans, any theory needs a mechanism to overlap the entropic gap between a highly diluted prebiotic ocean to a system that can kinetically sustain chemical reactions. A concentration increase of nucleotides exceeding $10^8$ fold has been demonstrated for porous mineral precipitates that layer the margines of hydrothermal vents, solving one of the most important open questions for the emergence of life, the so called *concentration problem* (Baaske et al., 2007).
Only "mild conditions" can allow the emergence of organization. The first example of this statement are the off-axis hydrothermal vents which created the first place for the emergence of life (Russell and Hall, 1997). The so called "black smokers" had a too high temperature to allow the emergence any form of life (Jupp and Schultz, 2004). In analogy, the proposed RNA - world could better undergo polymerization and replication of primitive nucleotides in a hydrothermal vent rather than by interaction with UV rays in a "primordial soup" (Baaske et al., 2007).
The Earth's state far from thermodynamic equilibrium is strongly related to the presence of life. The emergence of life allowed the use of more degrees of freedom associated to geological and atmospheric cycles, and consequently the generation of more free energy from the same initial energy sources. Lovelock (1965; 1975) noted that the Earth atmosphere is maintained far from thermodynamic equilibrium in contrast to its planetary neighbors and proposed the extent of disequilibrium as an indication of planet's habitability (Hitchcock and Lovelock, 1967).

There are important evidences that life affected geochemical cycles (Rosing et al., 2006. Dyke et al., 2010). Biotic activity altered and still alters the rates of geochemical reactions, accelerating the rates of silicate rock weathering by one to three orders of magnitude, thereby altering the geological carbon cycle and planetary habitability (Schwartzman & Volk, 1989. Berner, 1997).

The evolution of the sedimentary cycle, driven by atmosphere - water - geosphere interactions, gave a set of porous precipitates, which led to the confining phenomena needed for the emergence of life. The latter also changed the surface characteristics of the Earth, as reflected in the frequency of present topographic properties (Dietrich and Tayler Perron, 2006). Thus, it is always important to consider the interactions between Earth's geological evolution and life's colonization of the planet as co - evolving (Grenfell et al., 2010. Lammer et al., 2010).

Earth shows, as a far from thermodynamic equilibrium system, a continuous organization of the interplay between thermal and chemical processes, allowing the emergence of new structures. Every step that drove to the emergence of life saw the interplay between convective mass transfer of heat and chemical gradients. Thermal gradients can establish chemical potential contrasts. Structures connected with the emergence of life evolved, increasing their reactivity, through a catalytic potential. In this paper, a simplified thermodynamic description of the production of chemical disequilibrium by these processes is given. We begin with a description of the passage from hydrothermal heat to electrochemical energy, with a first order calculation of the energy flow. Then, a general mechanism of chemical free energy generation and the rise of chemiosmosis in confined, inorganic matrices is presented. After a short overview of the possible matrices and their catalytic properties for the rise of molecular complexity and the emergence of life, we propose a first order evaluation of the thermodynamics of a simplified RNA - clay system.

Figure 2: The general mechanism of chemical free energy generation.

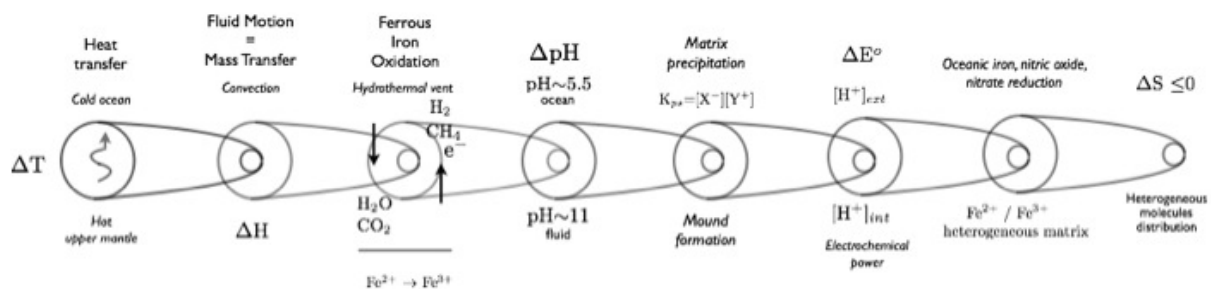

## 2. The general mechanism of Chemical Free Energy Generation

The different steps involved in generating chemical free energy from heat flow at the Earth's crust-ocean interface are shown in Figure 2. In the transfer of free energy from one process to the following, from the geothermal heat flux to the generation of chemical disequilibrium, thermodynamic inefficiencies result in successively smaller rates of free energy transfer.

Since the continuous generation of geochemical free energy is a basic requirement for the emergence and evolution of life, its greater capture would seem to be a key issue in order to develop and maintain metabolisms. This process is argued to happen through chemiosmosis, the process by which free energy is harvested from an electrochemical disequilibrium (Ducluzeau et al., 2009). Thermodynamic inefficiencies and free energy transfers are explained in more detail in the following.

When the upper, reduced and hot mantle comes in contact with water permeated in the oceanic crust, the difference in temperature generates the kinetic free energy associated with fluid motion. Porous rocks, ocean and upper mantle have finite thermal capacities, allowing heat losses during its transfer to the fluid. Hence, the maximum efficiency for this process is much lower than the Carnot one. As a general statement, thermodynamic boundary conditions allow only for a fraction of the heat flux to be converted into the power to drive fluid motion.

The convective motion at the crust-ocean interface allows for the process of serpentinization, in which the rising, hot fluid is rich in reduced molecules (mainly $H_2$ and $CH_4$ by reduction of $H_2O$ and $CO_2$ through the oxidation of ferrous iron in olivine) which exhaled through the surface of the oceanic crust (Martin and Russell, 2003), which then relates the kinetic energy of the fluid motion to the generation of chemical free energy associated with the redox gradient between the rising fluid and the oxidizing ocean water. This kind of hydrothermal vent is placed faraway from a magmatic chamber (off - axis), allowing an outflowing fluid temperature not too high for any ancient living form (≤367 K) (Charlou et al., in press). The high temperature and basicity of the fluid pH (~11) led to the precipitation of new mineral formations: hydrothermal mounds, in and through which the fluid continued to outflow. A redox front formed, contributing to reduce $CO_2$, and the oceanic nitrate and $Fe^{3+}$. A mix of $Fe^{3+}$ and $Fe^{2+}$ contributed to the formation of an heterogeneous matrix (see Section 4), followed by adsorption and interaction of reduced organic molecules.

Boundaries are thermodynamically extremely important. The natural precipitation of matrices facilitate the formation of new boundaries. From the point of view of molecular interactions, the presence of a boundary affects the diffusive properties, enhancing the efficient use of free energy and molecular selectivity. Further, if an heterogeneous adsorbing matrix is built, concentration gradients of adsorbed molecules rise (Marin et al., 2009). Molecular gradients are maintained for a longer time due to compartments, decreasing the entropy production, $\sigma = \frac{dS}{dt}$. As a result, any attenuation of external gradients can be used to induce internal gradients, actually storing free energy, easily available for the synthesis of more complex organic compounds. A gradient from pH~10 to pH~8 could provide the chemical energy needed for a phase change from lipid micelle to vesicles (Russell 2003. Hanczyc et al., 2003). The continuous flow of reduced chemicals and alkaline, hot fluids in porous heterogeneous matrices on the side of hydrothermal vents, represented a dissipative structure which provided a basis for the rise of more complex molecules and allowed an initial randomization of matter and energy states (Prigogine 1978. Wicken 1978; 1980).

As a result of the chemical disequilibrium, reduced compounds confined in small matrices gave rise to a process known as chemiosmosis (Figure 1) (Russell et al., 1993; 1994. Russell and Hall, 1997. Lane et al., 2010). The alkaline hydrothermal fluids permeating the margins of the mound promoted a natural proton gradient as the mildly acidic early ocean water is met, so inducing a natural proton motive force, a process considered necessary for carbon and energy metabolism in the first chemotrophs, due to thermodynamic constrains (Lane et al., 2010). A proton motive force, in a general definition, is a $[H^+]$ and electric potential gradient that stores energy (in chemical bonds). Nowadays, cells use a proton motive force starting from an internal chemical energy source to synthesize ATP. A strong hypothesis invokes this natural proton motive force to take part in the synthesis of organic molecules which also supports the idea that for proto - metabolism the first main electron donor was hydrothermal $H_2$ and the main electron acceptor was $CO_2$ (Russell and Hall, 1997).

This situation could produce a range of different molecular monomers, and the free energy due to the proton motive force allowed the synthesis of reduced carbon and nitrogen compounds. Hence, the presence of an high number of monomers and a good linkage mechanism among them was the start of molecular evolution (Pulselli et al., 2009). Porous matrices provided a means to concentrate newly synthesized molecules, thereby increasing the chance of forming oligomers (Baaske et al., 2007); the temperature gradients inside the hydrothermal vent allowed "optimum zones" of partial reactions in different regions of the vent (e.g. monomer synthesis in the hotter, oligomerisation in the colder parts). In addition, compartmentalization processes in a catalytic matrix can open exponential replication ways for molecular evolution.

## 3. Initial energy supply: The Vent System

From a geological point of view, thermal gradients are the most abundant dissipative systems on the early Earth (Baaske et al., 2007). The off-axis vents are located several kilometers away from the spreading zone of the ocean ridge. Their waters do not come into close contact with the magma chamber, allowing the possibility for the emergence of life (Martin and Russell, 2003). Hydrothermal vents add together three important properties:
- the presence of small pores;
- constant fluid flux with a constant supply of negative charged entities;
- the presence of mild conditions and the influence of pH, redox and temperature gradients.

Alkaline hydrothermal fluids favor both phosphate and amine chemistry, and promote a proton - motive force across membranes, as required by cells (Russell 2003. Russell and Arndt, 2005. Russell and Kanik, 2010. Lowell and Rona, 2002).

The basis for chemical free energy generation is the convective motion of heated fluid. We now set up a simple model to illustrate that for a given heating rate, a maximum conversion rate to kinetic energy exists that can subsequently be used to drive other forms of free energy generation. The heating rate $Q_{serp}$ that the upwelling fluid receives from serpentinization can be calculated as:

$$1) \quad Q_{serp} = L \cdot M$$

where $L$ is the latent heat of reaction and $M$ is the rate of serpentinization. Considering data from the vent called *Lost City* (Lowell and Rona, 2002), for each km² of reacting upper mantle, $L=2.5 \times 10^5$ J kg$^{-1}$ and $10 \leq M \leq 100$ kg s$^{-1}$, we obtain that

$$2) \quad Q_{serp} = 2.5 \times 10^6 - 2.5 \times 10^7 W km^{-2}$$

A basic heat coming from the magma chambers (usually 15 km far away) of $Q_{heat} \sim 5 \cdot 10^5$ W is also added (data for *Lost City*). The upwelling fluid enters the ocean with $T_{fluid} \sim 367$ K (Charlou et al., in press). Hence, a difference in temperature can precipitate part of the effluent compounds, such as carbonates, hydroxides, siliceous gels and iron sulphides, which give rise to the hydrothermal mounds and the future catalytic chambers for complex organic compounds synthesis (see Section 4).
At a steady state, the energy balance inside the hydrothermal vent can be described as

$$3)\quad c\frac{dT_{serp}}{dt} = Q_{heat} + Q_{serp} - Q_{conv} - Q_{cond} = 0$$

where $Q_{conv}$ and $Q_{cond}$ are the convective and conductive heat fluxes respectively, and $T_{serp}$ is the temperature at which serpentinization occurs, in Kelvin, and $c$ is the thermal capacity of the oceanic crust. The conductive heat flux can be expressed as

$$4)\quad Q_{cond} = k_{cond} \cdot (T_{serp} - T_{ocean})$$

where $k_{cond}$ is the conductivity constant of the oceanic crust, $T_{ocean}$ is the temperature of the ocean far away from the vent; then, the entropy production by convection, $\sigma_{conv}$ (expressed in J K$^{-1}$ s$^{-1}$), is:

$$5)\quad \sigma_{conv} = Q_{conv} \cdot \left(\frac{1}{T_{ocean}} - \frac{1}{T_{serp}}\right) = Q_{conv} \cdot \frac{(T_{serp} - T_{ocean})}{(T_{serp} \cdot T_{ocean})}$$

Using the Eq. 3 and Eq. 4, we can write:

$$6)\quad \sigma_{conv} = Q_{conv} \cdot \frac{Q_{cond}}{k_{cond} \cdot (T_{serp} \cdot T_{ocean})} = Q_{conv} \cdot \frac{(Q_{heat} + P_{serp} - Q_{conv})}{k_{cond} \cdot (T_{serp} \cdot T_{ocean})}$$

Since this expression for entropy production by convection is a negative quadratic form of $Q_{conv}$, $\sigma_{conv}$ has a maximum value. Since entropy is produced by frictional dissipation ($\sigma_{conv} \approx \frac{D}{T}$ with $D$ being the rate of dissipation), and frictional dissipation balances generation of motion in steady state, this state of Maximum Entropy Production (MEP, Ozawa et al. 2003, Kleidon 2009) corresponds to one in which maximum kinetic energy is generated. Such maximum intensity of fluid motion should correspond to maximum transport of electrons, and therefore to a maximization of electrochemical power generation in the vent system. Here, we only use this example to point out the fundamental thermodynamic limits of how much free energy can be generated from a geothermal heat flux. Future work would be needed here to quantify the resulting generation rate of geochemical free energy.

Figure 3: Energetic scheme of the Serpentinization/Electrochemical Potential system.

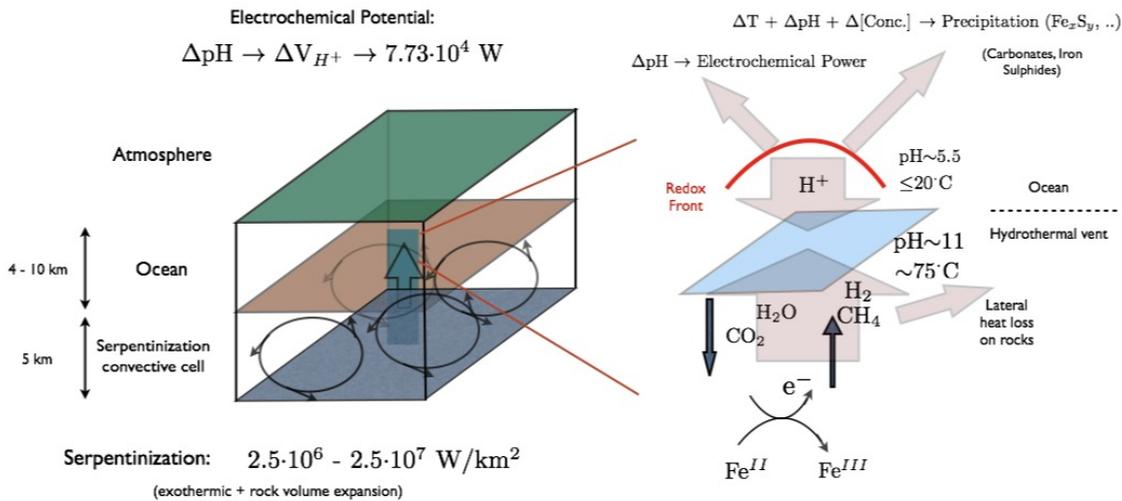

The Archaean ocean was weakly acidic (pH~5.5). A pH gradient could produce an Electrochemical Potential, according to the Nernst formulation. Since the operating ions are $H^+$, here we have:

$$7)\quad E_{ion,H^+} = \frac{R \cdot T}{z \cdot F} ln\frac{[H^+]_{external}}{[H^+]_{internal}}$$

where $R$ is the Boltzmann gas constant, $z$ the charge of the considered ion, $F$ the Faraday constant, $[H^+]_{external}$ and $[H^+]_{internal}$ are the oceanic and flux concentrations of $H^+$ in to the mound, respectively. This results in an electrochemical potential of 320.6 mV. Considering the reaction:

$$8)\quad H_2 \to 2H^+ + 2e^-$$

and that each electron has a charge of 1 C and an energy of 1eV = $1.602 \times 10^{-19}$ J, an energy of $1.03 \times 10^{-19}$ J/molecule or $6.19 \times 10^4$ J mol$^{-1}$ is obtained. Hydrogen production by serpentinization process is 0.5 mol $L_{olivine}^{-1}$ (Lowell and Rona, 2002), which means 1.25 mol$_{H2}$ s$^{-1}$ (in *Lost City*). Hence, the electrochemical power produced is $7.73 \times 10^4$ W for each square km of serpentinized upper mantle.

The complete Serpentinization / Electrochemical Potential system is summarized in Figure 3. Since $[H^+]$ in the outgoing fluid depends on $M$, and hence on $T_{serp}$ and $T_{ocean}$, the former is related to the entropy maximization in the hydrothermal vent.

## 4. Involved Matrices: The Catalytic Chambers

### 4.1 Early compartments

Hydrothermal vents constantly pumped alkaline warm solution into confining porous mound of freshly precipitated carbonates, clays, hydroxides, such as $Mg(OH)_2$, and in places on the early Earth, iron-nickel sulfides, with a wide presence of catalytic transition metals (Ni, Co, Mn, W, Zn, Mo), into a cool and acidulous ocean. Hence, hydrothermal mounds are composed of highly porous precipitates like aragonite ($CaCO_3$). They affected also the thermodynamics and chemistry of the enclosed environment: brucite ($Mg(OH)_2$) can stabilize pH at ~9.8, allowing a first peptide cycle (Huber and Wächtershäuser, 1998. Russell 2003.

Milner-White and Russell, 2008).

It is important to underline the precipitation of monosulfides such as FeS (which, on sulfidation gives pyrite, $FeS_2$) and sphalerite, ZnS. Their presence emphasizes the redox heterogeneity of the newly formed matrix, and is also important for their affinity to organophosphates, cyanide, amines and formaldehyde (Russell, 1994. Martin and Russell, 2003). In fact, at the ocean temperature and conditions near hydrothermal mounds, the reaction of oceanic $Fe^{2+}$ with $H_2S$ produced monosulfides, FeS, that could precipitate as metastable metal - sulfide gels. Nickel is usually an ancillary of iron in these structures, which have been addressed as the place where life emerged (Russell et al, 2005).

The strong disequilibrium given by the contemporaneous presence of hydrothermal $H_2$ and $CO_2$ dissolved in ancient oceans was not sufficient to overcome the reactions' kinetic barriers. Metal ions incorporated in the precipitated matrices had catalyzed the interaction of molecules, lowering their activation energy.

The most important metal complexes for these processes were mackinawite ([Fe>>Ni]S) and greigite ($NiS_2[Fe_4S_4]S_2Fe$), dispersed within the hydrothermal mound (Russell and Hall, 2006. Russell et al., 2008). This mound, effectively a catalytic flow reactor, supported strong redox and pH gradients and the compartments comprising the mound acted as suitable reaction sites. Hydrodynamic pressure could coat the internal part of these Fe - S cavities with organic molecules formed therein with high affinity with the matrix; peptides and other simple polymers could hence act as the first *organic* membrane, which still needed, however, iron-sulfur complexes as a source, and for the transfer of electrons. A comprehensive discussion of these metal clusters is given by Milner - White and Russell (2005, 2008) and Russell at al. (2005).

The early compartments so formed then assumed the role of catalytic chambers in which the concentration of molecules increased and reactions were catalyzed (Baaske et al., 2007. Russell et al., 2008). The metal - sulphur clusters comprising these minerals interacted catalytically with solved molecules of the alkaline fluids. The chelation of smaller metal - sulphur cluster by organic molecules allowed dissolution in water of these new catalytic molecules (Russell and Hall, 2001; Milner - White and Russell, 2008; Russell and Kanik, 2010).

The hydrothermal vent conditions also induce mineralogical, textural, geochemical and morphological patterns, resulting from the interplay of reaction and diffusion kinetics (Hopkinson et al., 1998. Ortoleva et al., 1987). These represent inorganic self - organized mechanisms of precipitate patterning (involving ferric iron oxides and hydroxides) due to far from equilibrium crystallization at redox and pH fronts.

**4.2 Clays**

In Section 2 we outlined one of the most important theories to explain the rise of molecular complexity (Wicken 1978. Pulselli et al., 2009). From a more practical point of view, the creation of more complex molecules is generally obtained by dehydration. Hence, it is difficult to imagine an origin by random collisions in the presence of a high concentration of water (Gallori et al., 2006. Pace 1991). For example, this is particular evident for RNA molecules, which bear an 2' - OH group susceptible to hydrolysis. It has long been proposed that surface chemistry on clays was involved in both hydrolysis and polymerizations (Ferris, 2002). In 1951, J. D. Bernal suggested that clay minerals could have bound organic molecules from the surrounding water, enhancing their concentration and protecting them against high temperatures and solar radiation. Numerous experimental observations have confirmed this original hypothesis in recent years (Ferris 2002. Smith 1998). For example, DNA originating from dead or living cells can persist for a long time in the environment, without losing its biological activity, as a result of its association with clay minerals (Gallori et al., 1994). In addition, clay particles with high surface charge, can catalyze the assembly of lipid vesicles in water (Hanczyc et al., 2003).

These experimental observations allowed the formulation of the RNA - Clay - world theory (Franchi et al., 2005). Pores in, for example, feldspars or iron sulfides offered themselves as natural staging posts for the evolution of replicators in compartmentalized, distinct groups and thus allowed for the coexistence of many more different types of metabolically cooperating replicators than flat mineral surfaces do (Koonin and Martin, 2005; Branciamore et al., 2009).

From a bottom - up point of view, we observe that hydrothermal alteration and deposition can result in clays and zeolites with their corresponding rock/water contact surfaces (Bonatti et al., 1983; Marteinsson et al., 2001; Hazen et al., 2008). The resulting increase in reactivity then parallels the increase of compartmentalization needed for life's emergence (Franchi et al., 2003. Franchi and Gallori, 2005. Segré et al., 2001).

**4.3 Proposed systems**

In order to characterize the behaviour of the considered reaction / matrix systems, some experiments are here proposed. The aim is to fill a lack between proposed thermodynamic theories and "real" chemistry, and to advance more simple and manageable systems.

A general basis is to consider dissipative structures (Prigogine 1967; 1978), which manifest both spatial and temporal periodicity, but cannot be described by any known potential function and do not show universal tendency toward entropy production (Rossi et al., 2008). If a self - assembled structure is combined with them, the system can show self - organization (Yamaguchi et al., 2004; 2005). One of the most studied chemical dissipative system is the Belousov - Zhabotinsky (BZ) reaction, which shows evolutive peculiarities of spatio - temporal manifestations (Belousov 1958; Rossi et al., 2008); the involved symmetry - breaking processes are interesting in the study of life emergence and evolution, expecially if BZ is combined with several kind of matrices. In fact, its behaviour with other biomimetic compartments shows very interesting patterns and has been studied in order to understand reaction - diffusion and reaction - diffusion - convection interactions (Magnani et al, 2004; Murray 2002; Rossi et al., 2008; Turing 1952; Vanag 2004). Further, chemical oscillators like BZ play a significant role in the understanding and modeling of several rhythmic manifestations of Life (Goldbeter 1996. Hess and Boiteux, 1971. Larter 2003. Müller and Hauser, 2000. Shanks 2001).

Proposed systems are shown in the bottom part of Table 1. Titanium oxide has an important catalytic role and oxidation activity associated to complex organic molecules (Saladino et al., 2003. Peteline and Yusfin, 1997). Beyond the already cited qualities of clays, their strong interaction with ferroin ($Fe^{2+}$(o-phen)$_3$ , o-phen being ortho-phenantroline) and other iron - bearing complexes (Ferreiro and de Bussetti, 2005. Yamagishi 1982) makes possible the interaction with BZ reaction a good experimental system for the study of chemical patterns and dissipative processes in this matrix.

Table 1: Matrices discussed in this paper and, in the bottom part, proposed experimental matrix/reaction systems, useful to study the rise of a self - sustained proto – metabolism.

| Compartment | Fluxes | Main cause | Main effect | Structure | Catalyst involved |
|---|---|---|---|---|---|
| Hydrothermal mound | hot, reduced fluid | Serpentinization Vent | pH, T gradients Reduced compounds $PO_4^{3-}$ chemistry Acetate | $Fe_xS_y$ sol-gel $CO_3^{2-}$ + (Fe, Ni)S Clays | Fe Proton - motive force |
| RNA - Clays | Aqueous fluid | Sedimentary cycle | Nucleotides polymers | Clays | Metals (Fe, Mn) |
| Proposed system | Reactant(s) | | Structure | Catalyst involved | |
| $TiO_2$ - BZ | Short carboxylic acid | | $TiO_2$ layers | Ti, OH | |
| BZ - Clays | Short carboxylic acid | | Clay layers | Metal-chelates, OH | |

## 5. Looking for an entropic threshold: application to the RNA-Clays world

Hydrothermal vents provided the sustained source of chemical energy that gave rise to several exergonic, spontaneous and inorganically catalyzed reactions (Martin et al, 2008). Subsequently, the presence of chemosynthetic pathways independent from vent electrochemical gradients (but still linked to their reduced organic flux) evolved more complex molecules (Lane et al., 2010). One of the most important molecules is an RNA - precursor which could have had auto - catalytic properties (Franchi and Gallori, 2005; Yarus, 2010).

It is nowadays well known that nucleotides, poly - nucleotides and RNA molecules can adsorb on clays surface (Branciamore et al., 2009. Franchi and Gallori, 2005. Franchi et al, 2003). Hanczyc et al. (2003) and Huber and Wächtershäuser (1997) demonstrated the emergence of RNA - bearing vesicles and the possibility of a primordial peptide cycle, both in alkaline conditions (Russell 2003).

In this section the intereaction of RNA with clays is taken into account, due to its importance in the Emergence of Life as an autocatalytic - inorganic surface - mediated system. A calculation of the entropy production of a simplified RNA - clay system, as a first application of the theory is shown below.

### 5.1 Methods

Consider two systems, one with a certain volume $V$, and a second system with the same total volume divided in $k$ compartments each of them with their own volume $V_k$. The chemical species $i$ has a molar fraction $\chi_i = \frac{n_i}{\Sigma_i n_i}$ in the first system, and $\chi_{i,k} = \frac{n_{i,k}}{\Sigma_{i,k} n_{i,k}}$ in each volume $V_k$. Marin et al. (2009) demonstrated that the entropy difference between a unique volume system and a compartmentalized system is

$$9) \quad \Delta S_i = R n_i \Sigma_k \left( \frac{\chi_{i,k}}{\chi_i} \frac{V_k}{V} \right) ln \frac{\chi_i}{\chi_{i,k}}$$

It follows that if the component $i$ is equally distributed between all compartments $k$, then and $\Delta S_i = 0$. However, if a system evolves with a gradient of the species $i$ between the $\chi_i = \chi_{i,k}$ compartments, then the system has lower entropy than a non-compartmentalized one (Figure 4). The system can use the depletion of this concentration gradient to obtain free energy.

Figure 4: Compartmentalization of a same volume can decrease entropy and create internal gradients in a system, if an heterogeneous adsorption for one or more compounds is present among the internal compartments.

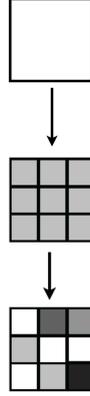

Following Kondepudi and Prigogine (2003) the entropy production due to chemical reactions can be calculated by:

$$10) \quad \left(\frac{1}{V}\right)\left(\frac{dS}{dt}\right) = \left(\frac{1}{V}\right)\left(\frac{A}{T}\right)\left(\frac{d\zeta}{dt}\right)$$

where $V$ is the total volume of the system, expressed in liters, $A$ is the chemical affinity of each reaction, $T$ the reaction temperature in Kelvin and $\zeta$ is the reaction coordinate. For a simple step reaction, this can be written as:

$$11) \quad \left(\frac{1}{V}\right)\left(\frac{dS}{dt}\right) = R(R_f - R_r)ln\left(\frac{R_f}{R_r}\right)$$

where $R$ is the Boltzmann gas constant and reaction rates $R_f$, $R_r$ can be assumed by reaction formula (still, considering a simple reaction step).

In the following, we consider a system in which the starting concentration of a ~3400 nucleotides ribosomal RNA is $[rRNA]_i = 1.29 \times 10^{-9}$ M dissolved in a unique reactor of 1 L. A general and fulfilling reference for this system is Gallori et al. (2006). As shown in Franchi et al. (1999), the adsorption of ribosomal rRNA on montmorillonite and on kaolinite is 5.88 µg/mL and 4.38 µg/mL, respectively. Then, we consider a hammerhead ribozyme, with an initial concentration $[hRNA]_i \sim 1 \times 10^{-13}$ M (Biondi et al., 2007), which undergoes a self-cleavage reaction, enhanced in particular conditions by its interaction with the clay matrix. Finally, we consider the formation of boundaries inside the system, increasing its compartmentalization.

**5.2 RNA adsorption on clays**
A fulfilling description of the RNA - clay interaction from a kinetic point of view can be found in Franchi et al. (1999). Considering the adsorption reaction as a chemical process, as

$$12) \quad rRNA \rightarrow (rRNA - Clay)_{complex}$$

at the equilibrium we have $K_{eq} = \frac{[rRNA-Clay_{complex}]}{[rRNA]}$. The entropy to form the adsorbed complex can be calculated as $\Delta S^f = -R \cdot ln(K_{eq})$. Using data shown above, we obtain an entropy of 7.79 J K$^{-1}$ mol$^{-1}$ for the adsorption on montmorillonite, and 10.26 J K$^{-1}$ mol$^{-1}$ on Kaolinite.

**5.3 Clay catalytic power**
In order to have an example of the entropy production due to a reaction occurring on a catalytic surface, we consider one of the most important process due to hammerhead – RNA; the self - cleavage reaction. Although this process is governed by a Michaelis - Menten kinetics, from the entropic point of view it can be simply described as a first order reaction:

$$13) \quad hRNA \underset{k_r}{\overset{k_f}{\rightleftarrows}} RNA_1 + RNA_2$$

where *hRNA* is the starting chain, and *RNA$_1$*, *RNA$_2$* are the two pieces of the reacted chain. As shown in Biondi et al. (2007), the reaction without clays has a first order rate constant of $k_{obs}$=0.027 min$^{-1}$ while in the presence of clays $k_{obs}$=0.343 min$^{-1}$. In the second, the reaction is strongly overbalanced towards products, hence $k_f$=$k_{obs}$ and $k_r$ can be assumed at a very low value (1x10$^{-20}$ L mol$^{-1}$ s$^{-1}$).

For reaction in Eq. 13, we have:
$$14) \quad R_f = R_{obs} = k_{obs}[hRNA]$$

$$15) \quad R_r = k_r[RNA_1][RNA_2]$$

Simple simulation of reaction in Eq. 13 showed that chemical equilibrium can be reached after 10$^4$ s (~ 2h 50 min). In this time span the entropy produced in a reactor with volume of 1 L is 9.66x10$^2$ J K$^{-1}$ mol$^{-1}$ without montmorillonite, whereas that produced with montmorillonite is 1.39x10$^3$ J K$^{-1}$ mol$^{-1}$. In the considered time span, RNA - cleavage reaction is enhanced, as we can see in Figure 5.

Figure 5: Simulated consumption of hRNA by self - cleavage, as supposed in Reaction in Eq. 13. Left: without montmorillonite; right: with montmorillonite. Simulation obtained with COPASI package.

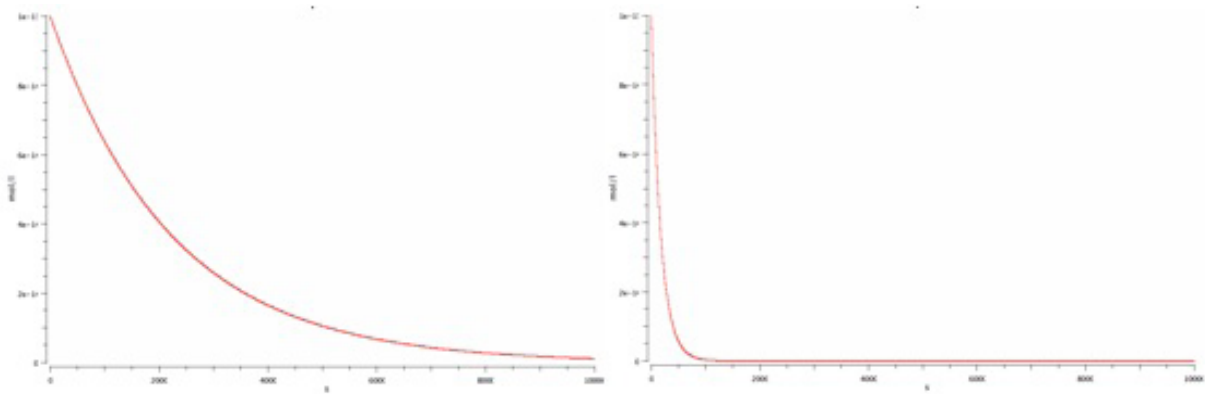

**5.4 Heterogeneous confinement**
Considering Eq. 9, a simple example of net entropy decrease due to heterogeneous compartmentalization is now proposed. Montmorillonite is considered dissolved in the reactor, which evolves internal sub - compartments, still with montmorillonite or kaolinite dissolved. Using data discussed in Section 5.2, the constitution of a new setup with two, smaller and differently adsorbing compartments can be calculated by Eq. 9 in -1.15x10$^{-9}$ J K$^{-1}$ L$^{-1}$, or -1.15x10$^4$ J K$^{-1}$ mol$^{-1}$. A more complex conformation can be formed by three sub - compartments, one with montmorillonite ($V_l$=0.45 L), the second with kaolinite ($V_l$=0.28 L), and the third with no clay ($V_l$=0.27 L); in this case, the entropy decrease is -1.13x10$^{-9}$ J K$^{-1}$ L$^{-1}$, or -1.13x10$^4$ J K$^{-1}$ mol$^{-1}$.

## 6. Discussion

In this paper we showed how chemical potential gradients result in a mass transfer that is often selective. First, our schematic first order entropic model of the hydrothermal vents showed that it is possible to calcualte Maximum Entropy Production for this system, from which we can define a upper limit for the serpentinization process and so for the maintenance a continuous redox front. In those conditions, the systems seems to reach a steady state.

The natural contemporaneous presence of a redox front, and heat flow, different catalytic centers in a confining matrix acted as a catalytic chamber for the emergence of more complex organic compounds and redox chains (Martin and Russell, 2003. Russell and Hall, 1997. Lane et al., 2010). One of the perspectives after the present paper is to compute a complete modeling of the hydrothermal vent system.

The proposed, schematic theoretical approach for the production of chemical free energy has been applied to an important model in the emergence of life, the RNA - clay system. Our first order calculation shows that the RNA - clay adsorption and a clay - kinetically favored RNA reaction, the self - cleavage, both increase the entropy produced by the system; however, the formation of sub - compartments, with an heterogeneous distribution of adsorbing matrix, decreases the total entropy produced, with a final entropy balance of $\Delta S \sim -10^4$ J K$^{-1}$ mol$^{-1}$. Results are summarized in Tab. 2.

Table 2: RNA-clay preliminary entropic study.

| Process | $\Delta S$ / (J K$^{-1}$ mol$^{-1}$) |
|---|---|
| Adsorption on montmorillonite | 7.79 |
| Adsorption on kaolinite | 10.26 |
| Self - Cleavage w/o montmorillonite | $9.66 \times 10^2$ |
| Self - Cleavage with montmorillonite | $1.39 \times 10^3$ |
| Heterogeneous compartments formation | $-1.15 \times 10^4$ |

Heterogeneous compartmentalization must take place before there can be any autocatalytic reaction, and to afford a source of chemical free energy. This energy supported RNA self - cleavage on clays increases the entropy produced, although the kinetics of the reaction is favored. Hence, the RNA - cleavage reaction in the presence of clays produces more entropy in a shorter time (see the time needed to reach the equilibrium, in Fig. 5). This means that the interaction of the important molecule, RNA, with clays increases entropy production associated with RNA self - replication; this theoretically implicates an exponential increase of the entropy production, until the matrix or the considered system is saturated or the self - cleavage reaction stops for other causes. The final result is the stabilization of the system in a further-from-equilibrium state, able to produce more free energy by the attenuation of concentration gradients.

The proposed RNA - clay modeling is only a starting point for the application of the far-from-equilibrium thermodynamics approach in this research field. To complete a compartmentalization - protometabolism model of the RNA - clay - honeycomb (Branciamore et al., 2009) is another task for the future.

To conclude, in this paper we proposed an entropic explanation for chemical free energy production. An important factor is the selectivity for a certain species by a sub-compartment of the system, which has been shown to be a tool to store free energy.

**Acknowledgements**

The authors kindly acknowledge James Dyke for fruitfull discussion and test revision, the Helmholtz Alliance for Planetary Evolution and Life, Germany, for financial research support, and the Editor of Journal of Cosmology – Special Issue for fruitfull suggestions.